\documentclass[twocolumn,aps,prl]{revtex4}
\usepackage{graphicx}

\begin{document}

\preprint{APS}
\title{On the analytic structure of Green's function for the Fano - Anderson model}
\author{E. Kogan}
\email{kogan@mail.biu.ac.il}
\affiliation{Jack and Pearl Resnick
Institute, Physics Department, Bar Ilan University, Ramat Gan 52900,
Israel}
\date{\today}

\begin{abstract}
We study analytic structure of the Green's function (GF) for the
exactly solvable Fano-Anderson model. We analyze the GF poles,
branch points and Riemann surface, and show how the  Fermi's Golden
Rule, valid in perturbative regime for not to large time, appears in
this context. The knowledge of analytic structure of the GF in
frequency representation opens opportunities for obtaining formulas
for the GF in time representation alternative to the standard one
using the spectral density.

\end{abstract}
\maketitle

The Fano-Anderson model \cite{fano}, which presents discrete level
coupled to continuum  is probably one of the simplest and best known
in quantum mechanics.  It is exactly solvable, and the solution of
the model in terms of Green's functions (GF) is presented in
\cite{mahan}. The spectral line intensity calculated in the paper by
Fano are conveniently represented using the Green's function
spectral density \cite{mahan}. When the problem is treated within
the quantum mechanics proper, one is typically interested in the
tunneling of the particle, initially localized at the discrete level
into continuum (see e.g. Cohen-Tannoudji et. al. \cite{cohen}). The
non-tunneling amplitude is standardly  expresses also through the
Green's function spectral density, so pragmatically speaking this is
what we need. However, in this paper we would like to see the
Green's function in a broader context, and study its analytical
structure as a function of frequency. We'll see that Green's
function (in frequency representation) is a multi valued function,
and on simple examples study its branch points, poles and Riemann
surface. In fact, we'll see that this study can be of practical
value also, because it opens new opportunities to connect  the
frequency representation of the Green's function (used in
calculation of the spectral line intensity) with the time
representation, necessary for the quantum mechanics proper.

To have everything at hand let us present a pedagogical derivation
of the non-tunneling amplitude. Our system consists of the continuum
band, the states bearing index $k$, and the discrete state $d$,
having energy $\epsilon$. The Hamiltonian of the problem is
\begin{eqnarray}
\label{ham}
H=\sum_k\omega_k\left|k\right>\left<k\right|
+\epsilon \left|d\right>\left<d\right|
+\sum_k\big(V_k\left|k\right>\left<d\right|
+h.c.\big),
\end{eqnarray}
where $\left|k\right>$ is a band state and $\left|d\right>$ is the
state localized at site $d$; h.c. stands for the Hermitian
conjugate. The wave-function  can be presented as
\begin{eqnarray}
\label{ab}
\psi(t)=g(t)\left|d\right>+\sum_{k} b(k,t)\left|k\right>,
\end{eqnarray}
with the initial conditions $g(0)=1$, $b(k,0)=0$. Notice that the
non-tunneling amplitude is just the appropriate GF in time
representation. Schroedinger Equation for the model considered takes
the form
\begin{eqnarray}
\label{dif}
i\frac{dg(t)}{dt}=\epsilon g(t)+\sum_{k}V_k^* b(k,t)\nonumber\\
i\frac{db(k,t)}{dt}=\omega_k b(k,t)+V_k g(t)
\end{eqnarray}

Making Fourier transformation (Im $\omega>0$)
\begin{eqnarray}
g(\omega)=\int_{0}^{\infty}g(t)e^{i\omega t}dt,
\end{eqnarray}
we obtain
\begin{eqnarray}
\label{eq}
-i+\omega g(\omega)&=&\epsilon g(\omega)+\sum_{k}V_k^* b(k,\omega)\nonumber\\
\omega b(k,\omega)&=&\omega_k b(k,\omega)+V_k g(\omega).
\end{eqnarray}
For the amplitude to find electron at the  discrete level,
straightforward algebra gives
\begin{eqnarray}
\label{int}
g(t)=\frac{1}{2\pi i}\int g(\omega)e^{-i\omega t}d\omega,
\end{eqnarray}
where
\begin{eqnarray}
\label{exact}
g(\omega)=\frac{1}{\omega-\epsilon-\Sigma(\omega)},\;
\end{eqnarray}
and
\begin{eqnarray}
\label{sigma}
\Sigma(\omega)=\sum_k\frac{|V_k|^2}{\omega-\omega_k}
\end{eqnarray}
The integration in Eq. (\ref{int}) is along any infinite straight
line parallel to real axis in the upper   half plane of the complex
$\omega$ plane. Notice that $g(\omega)$ is the  GF in frequency
representation. The quantity $\Sigma(\omega)$  is self-energy (or
mass operator).

For tunneling into continuum, the sum in Eq. (\ref{sigma})
should be considered as an integral, and Eq. (\ref{sigma}) takes the form
\begin{eqnarray}
\label{sigma3}
\Sigma(\omega)=\int_{E_b}^{E_t}
\frac{\Delta(E)}{\omega-E}dE,
\end{eqnarray}
where
\begin{eqnarray}
\Delta(E)=\sum_{k}
|V_k|^2\delta(E-\omega_k),
\end{eqnarray}
where and the limit of integration are the band bottom $E_b$ and the
top of the band $E_t$. We would like to calculate integral
(\ref{int}) closing the integration contour  by a semi-circle of an
infinite radius in the lower half-plane. Thus we need to continue
analytically the function $g(\omega)$ which was defined initially in
the upper half plane (excluding real axis) to the whole complex
plane. We can do it quite simply, by considering Eqs. (\ref{exact})
and (\ref{sigma3}) as defining propagator
 in the whole complex plane, save an interval of
real axis between the points $E_b$ and $E_t$, where Eq.
(\ref{sigma3}) is undetermined. (Propagator analytically continued
in such a way we'll call the standard propagator.) Thus the integral
is determined by the integral of the sides of the branch cut between
the points $E_b$ and $E_t$.
\begin{figure}
\includegraphics[angle=0,width=0.45\textwidth]{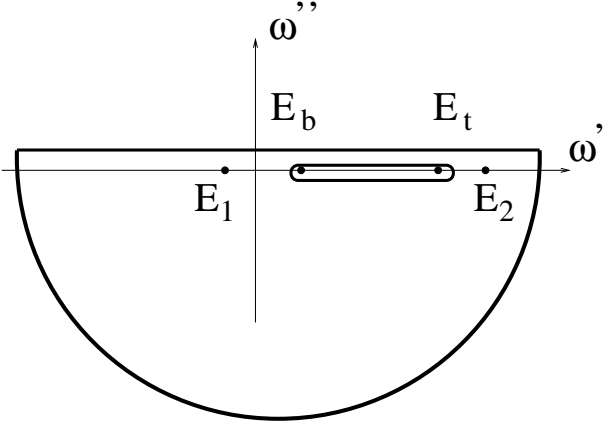}
\caption{\label{fig:cont} Contour used to evaluate integral
(\ref{int}). Radius of the arc goes to infinity.}
\end{figure}
The real part of the self-energy $\Sigma'$ is continuous across the
cut, and the imaginary part $\Sigma''$ changes sign
\begin{eqnarray}
-\Sigma''(E+is)=\Sigma''(E-is)=\pi\Delta(E)\quad s\to +0.
\end{eqnarray}
So the integral
along the branch cut is
\begin{eqnarray}
\label{cut}
I_{cut}=\int_{E_b}^{E_t}\frac{\Delta(E)e^{-iEt}dE}
{\left[E-\epsilon-\Sigma'(E)\right]^2+\pi^2{\Delta^2(E)}}.
\end{eqnarray}
Thus we have
\begin{eqnarray}
\label{ft0} g(t)=I_{cut}(t),
\end{eqnarray}
and the survival probability $p(t)$ is
\begin{eqnarray}
p(t)=|g(t)|^2.
\end{eqnarray}

In the perturbative regime
$|\Sigma'(\epsilon)|,|\Sigma''(\epsilon)|\ll
\epsilon-E_b,E_t-\epsilon$ the main contribution to the integral
(\ref{cut}) comes from the region $E\sim \epsilon$. Hence the
integral can be presented as
\begin{eqnarray}
\label{app}
I_{cut}=\int_{-\infty}^{\infty}\frac{\Delta(\epsilon)e^{-iEt}dE}
{(E-\epsilon-\Sigma'(\epsilon))^2+\pi^2{\Delta^2(\epsilon)}}
\end{eqnarray}
and easily calculated to give the well known Fermi's golden rule
(FGR)
\begin{eqnarray}
\label{fgr}
p(t)=e^{-t/\tau},
\end{eqnarray}
where
\begin{eqnarray}
1/\tau=2\pi\Delta(\epsilon).
\end{eqnarray}

However, even in the perturbative regime, the FGR has a limited
time-domain of applicability \cite{cohen}. For large $t$  the
survival probability  is determined by the contribution to the
integral (\ref{cut}) coming from the end points. This contribution
can be evaluated even without assuming that the coupling is
perturbative. Let $\Delta(E)\sim (E-E_b)^{\beta}$ $\beta >0$) near
the band bottom (the contribution from the other end point is
similar). Then  for large $t$
\begin{eqnarray}
\label{b}
 I^{(b)}_{cut}\sim t^{-(\beta+1)}.
\end{eqnarray}
The similar contribution comes from the top of the band.

If there can exist  poles of the propagator (\ref{exact}),  we
should add  the  residues to the integral (\ref{cut}). Thus we
obtain
\begin{eqnarray}
\label{ft}
g(t)=I_{cut}(t)+\sum_jR_j,
\end{eqnarray}
where the index $j$ enumerates all the real poles $E_j$ of the integrand, and
\begin{eqnarray}
\label{resid}
R_j=\frac{e^{-iE_jt}}{1-
\left.\frac{d\Sigma'}{dE}\right|_{E=E_j}}.
\end{eqnarray}
Notice, that the poles correspond to the energies of bound states
which can possibly occur for $E<E_b$ or $E>E_t$, and which are given
by the Equation
\begin{eqnarray}
E_j=\epsilon+\sum_{k} \frac{|V_k|^2}{E_j-\omega_k}.
\end{eqnarray}
If we take into account that normalized bound states are
\begin{eqnarray}
\left|E_j\right>=\frac{\left|d\right>+\sum_{k}
\frac{V_k}{E_j-\omega_k}\left|k\right>}{\left[1+\sum_{k}
\frac{|V_k|^2}{\left(E_j-\omega_k\right)^2}\right]^{1/2}},
\end{eqnarray}
then the residue can be easily interpreted as the amplitude of the
bound state in the initial state $\left|d\right>$, times the
evolution operator of the bound state times the amplitude of the
state $\left|d\right>$ in the bound state
\begin{eqnarray}
R_j=\left<d\right.\left|E_j\right>\left<E_j\right.\left|d\right>e^{-iE_jt}.
\end{eqnarray}
 If the propagator has one real pole at $E_1$, from Eq. (\ref{ft})
we see that the survival probability $p(t)\to |R_1|^2$ when $t\to
\infty$. If there are several poles, this equation gives Rabi
oscillations.

 Notice that Eq. (\ref{ft}) is just the well known
result \cite{mahan}
\begin{equation}
g(t)=\int_{-\infty}^{\infty}A(\omega)e^{-i\omega t}d\omega,
\end{equation}
where
\begin{equation}
\label{spectral}
A(\omega)=-\frac{1}{\pi} \text{Im}\left[g(E+is)\right]
\end{equation}
is the spectral density function. The first term in Eq. (\ref{ft})
is the contribution from the continuous spectrum, and the second
term is the contribution from the discrete states.

An indication that there is more in the GF than we have so far
discussed comes from the following fact: we could have obtained the
FGR in perturbative regime  directly from Eq. (\ref{int}), changing
exact Green function (\ref{exact}) to an approximate one, which may
be called the FGR propagator
\begin{eqnarray}
\label{fgrr}
g_{FGR}(\omega)=\frac{1}{\omega-\epsilon-\Sigma'(\epsilon)+i\pi\Delta(\epsilon)}.
\end{eqnarray}
 Thus approximated,
propagator has a simple pole $\omega=\epsilon+ \Sigma(\epsilon)$,
and the residue gives Eq. (\ref{fgr}). Thus, the  propagator which
is used to obtain Eq. (\ref{app})  and the FGR propagator have
totally different singularities, and still give the same survival
probability (in finite time interval).

The explanation is that the GF is  multi valued, and different
sheets of the function can be used to calculate integral
(\ref{int}). These ideas can be easily understood for  simple
examples. Consider a site coupled to a semi-infinite lattice
\cite{longhi}. The system is described by the tight-binding
Hamiltonian
\begin{eqnarray}
\label{ham2}
&&H=-\frac{1}{2}\sum_{n=1}^{\infty}\big(\left|n\right>\left<n+1\right|
+\left|n+1\right>\left<n\right|\big)
\nonumber\\
&&+\epsilon\left|d\right>\left<d\right|-V\big(\left|d\right>\left<1\right|
+\left|1\right>\left<d\right|\big),
\end{eqnarray}
where $\left|n\right>$ is the state localized at the $n$-th site of the
lattice. The band (lattice) states are described by the Hamiltonian
\begin{eqnarray}
H_0=-\sum_k\cos k\left|k\right>\left<k\right|,
\end{eqnarray}
where $\left|k\right>=\sqrt{2N}\sum_n\sin (kn)\;\left|n\right>$.
Hence we regain Hamiltonian (\ref{ham}) with $V_k=-\sqrt{2}V\sin k$.
After simple algebra we obtain (in the upper half plain)
\begin{eqnarray}
\label{sig}
\Sigma(\omega)=\Delta_0\left(\omega-\sqrt{\omega^2-1}\right),
\end{eqnarray}
where the square root is defined as having the phase $\pi/2$ just
above the real axis between $-1$ and $1$, and $\Delta_0=2V^2$. We
immediately see that the GF for this model is a double valued
function, the branch points being $+1$ and $-1$. The poles are given
by the equation
\begin{eqnarray}
\omega_{1,2}=\frac{\epsilon(1-\Delta_0)\pm\Delta_0\sqrt{\epsilon^2-1+2\Delta_0}}{1-2\Delta_0}.
\end{eqnarray}
One sheet has real poles for $\Delta_0\geq(\epsilon^2+1)/2$. For
$\Delta_0=(\epsilon^2-1)/2$ the GF has a second order pole at
$\omega=(\epsilon^2+1)/2\epsilon$. When $\Delta_0$ increases, this
second order pole is split into two first order poles, one going
right (we assume  $\epsilon>0$) and at $\Delta_0=1/2$ becoming a
pole at the infinity. For $\Delta_0>1/2$ this pole appears for
$\omega<-1$. The second  first order pole, when $\Delta_0$ increases
initially approaches the point $\omega=1$, and at a further increase
of $\Delta_0$ moves in the opposite direction and asymptotically
goes to infinity.

For $\Delta_0<(\epsilon^2+1)/2$  the second sheet  has two complex
poles of the first order. For $\Delta_0\ll 1$ the pole in the lower
half-plain is situated at $\epsilon -i\Delta_0\sqrt{1-\epsilon^2}$
and is just the FGR pole mentioned above. The poles position is
presented on Fig. 2.
\begin{figure}
\includegraphics[angle=0,width=0.45\textwidth]{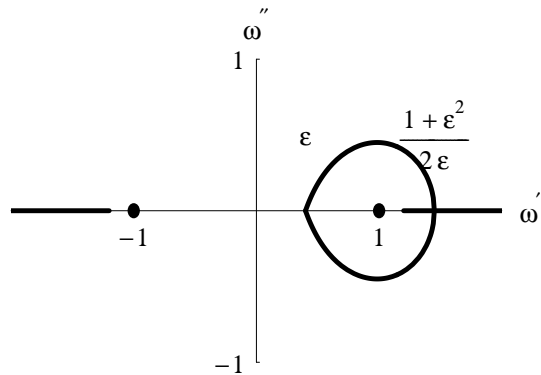}
\caption{Position of poles of the GF for the Hamiltonian
(\ref{ham2}) for different values of $\Delta$. The real poles appear
on the standard sheet, the complex poles on the second sheet.}
\end{figure}

In  general case the points $E_b$ and $E_t$ are the $\Sigma(\omega)$
branch points. The standard approach consists of making the cut
along the straight line between the branch points and  considering
only one sheet, thus making the analytic continuation into the lower
$\omega$ half-plain by continuing $\Sigma(\omega)$ along the curves
which circumvent the right branch point clockwise and the and the
left branch point anti-clockwise
 On the other hand, we
could use a different continuation, along the curve passing  through
the part of real axis between $E_b$ and $E_t$. In this case, Eq.
(\ref{fgrr}) which is valid in the perturbative regime near the
point $\omega=\epsilon$ in the upper half-plane is valid in lower
half-plain also, thus giving a pole in a different sheet of the
multivalued propagator.

This pole becomes important if  the analytical continuation of
$g(\omega)$ into the lower half-plain is done by making the cuts
from the branch points to infinity and continuing the function
between the cuts along the curves passing through the part of real
axis between $E_b$ and $E_t$, and outside as we did it previously.
This way to make an analytic continuation, and hence to calculate
the integral (\ref{int}) is presented on Fig. 3.
\begin{figure}
\includegraphics[angle=0,width=0.45\textwidth]{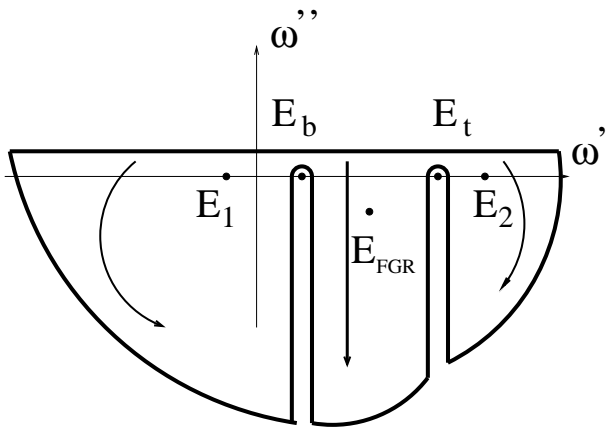}
\caption{\label{fig:cont2} Alternative way to analytically continue
the propagator into the lower half-plane. The arrows show the curves
of analytical continuation in between the cuts and outside.
 Radius of the big arc and length of the cuts go to infinity.}
\end{figure}
Due to the exponential decrease of the integrand on the vertical
line, in contrast to the oscillatory dependence on the real axis,
such analytic continuation is more convenient for the numerical
calculations for the large $t$ behavior of the non-decay amplitude.

The second example is defined by the equation
\begin{eqnarray}
\label{const} \Delta(E)=\Delta_0=\text{const}\qquad
\text{for}\;|E|\leq 1.
\end{eqnarray}
Thus we get
\begin{eqnarray}
\Sigma(\omega)=\Delta_0\log\left(\frac{\omega+1}{\omega-1}\right).
\end{eqnarray}
The Riemann surface has an infinite number of sheets. The standard
sheet is obtained by defining $\log$ as having the phase $-\pi$ just
above the real axis between $-1$ and $1$. This sheet always has  two
real poles, one for $\omega>1$, and the other for $\omega<-1$.

 For the pedagogic purposes let us
presents the results of numerical calculations for the model
considered.
 The time will be measured in units of the  FGR time $\tau$
\begin{eqnarray}
1/\tau=2\pi\Delta_0.
\end{eqnarray}
For the sake of definiteness we will chose $\epsilon =-.4$. For
$\Delta_0=.02$ (see Fig. 4) we observe  the FGR regime, say, up to
$t=9$.
\begin{figure}
\includegraphics[angle=0,width=0.45\textwidth]{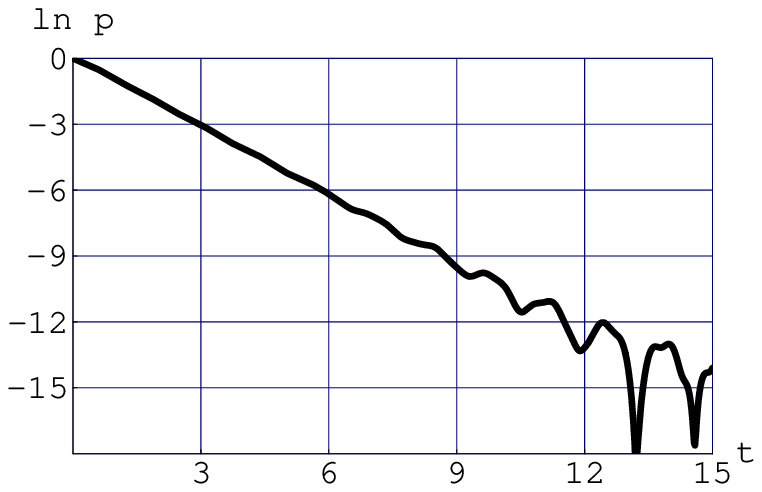}
\caption{Survival probability as a function of time  for $\Delta_0=.02$.}
\end{figure}
For $\Delta=.1$ (see Fig. 5)  the FGR regime is seen up to $t=3$.
\begin{figure}
\includegraphics[angle=0,width=0.45\textwidth]{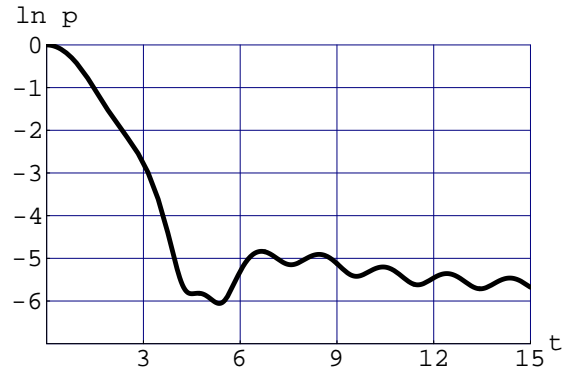}
\caption{Survival probability as a function of time for $\Delta_0=.1$.}
\end{figure}
For $\Delta_0=.2$ (see Fig. 6)  the FGR regime is absent.
\begin{figure}
\includegraphics[angle=0,width=0.45\textwidth]{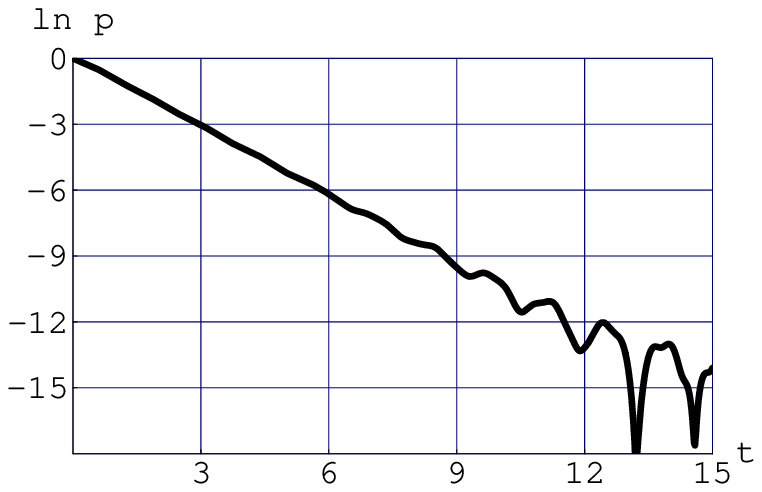}
\caption{Survival probability as a function of time for $\Delta_0=.2$.}
\end{figure}
The Rabi oscillations we see already at Fig.5 and still more vividly
at Fig.6.

\section{Dedication}
 With great pleasure I recall my meetings with I. Vagner
and our talks both about physics, and life in general.
In addition to being a very good physicist, he also was a very good educator,
investing a lot of time and effort in his pedagogical activities. So
I would like to dedicate this my modest contribution, mainly of pedagogical nature, to the memory of I. Vagner,
who's untimely death I deeply regret.

\end{document}